\documentclass[aps,prl,preprint,showpacs,groupedaddress]{revtex4}
\usepackage {amssymb}
\usepackage {amsmath}
\usepackage {graphicx}
\usepackage {longtable}

\begin{document}
\title{Ferroelastic fluctuations in high-temperature superconductors}

\author{Fedor V.Prigara}
\affiliation{Institute of Physics and Technology,
Russian Academy of Sciences,\\
21 Universitetskaya, Yaroslavl 150007, Russia}
\email{fprigara@recnti.uniyar.ac.ru}

\date{\today}

\begin{abstract}

It is shown that the specific heat jump at the superconducting transition
temperature in pnictide and cuprate superconductors is produced by
ferroelastic fluctuations just below the transition temperature. The
amplitude of the corresponding lattice distortion is estimated. It is shown
that a contribution from ferroelastic fluctuations to the specific heat jump
is also present in low-temperature superconductors.

\end{abstract}


\maketitle

Recently, it was shown [1] that the specific heat jump $\Delta
C_{P} $ at the superconducting transition temperature $T_{c} $ in
pnictide superconductors of $Ba\left( {Fe_{1 - x} Ni_{x}}
\right)_{2} As_{2} $ and $Ba\left( {Fe_{1 - x} Co_{x}} \right)_{2}
As_{2} $ series is proportional to $T_{c}^{3} $. Since the Debye
temperature $\theta _{D} $ in these compounds is weakly dependent
on chemical composition [2] (it is close to those of arsenic As),
this relation means that the specific heat jump $\Delta C_{P} $ at
the superconducting transition temperature $T_{c} $ is about 1\%
of the lattice heat capacity $C_{P} $ at this temperature,

\begin{equation}
\label{eq1}
\Delta C_{P} \approx 0.01C_{P} = 0.01 \times \left( {12/5} \right)\pi
^{4}Nk_{B} \left( {T_{c} /\theta _{D}}  \right)^{3},
\end{equation}

\noindent
where \textit{N} is the number of atoms and $k_{B} $ is the Boltzmann
constant.

The specific heat jump at $T_{c} $ in $La_{2} CuO_{4.093} $ [3] also obeys
the relation (\ref{eq1}). This relation is produced by ferroelastic fluctuations in
the superconducting phase just below the transition temperature [4] and can
be attributed to a relative change in the Debye temperature $\theta _{D} $
at a level of 0.3\%. The amplitude of the corresponding lattice distortion
$\delta = \Delta a/a$ (\textit{a} is the lattice parameter) is about $\delta
\cong 1.5 \times 10^{ - 3}$, if we assume that a relative change in the
sound velocity has the same order as a relative change in the lattice
parameter. This value of $\delta $ has an order of

\begin{equation}
\label{eq2}
\delta \cong a_{0} /d_{c} ,
\end{equation}

\noindent
where $a_{0} \approx 0.45nm$ has an order of the lattice parameter and
$d_{c} \approx 180nm$ is the size of a crystalline domain [5].

The maximum elastic strain in a crystalline solid and, hence, the
ratio of the tensile strength $\sigma _{s} $ to the Young modulus
\textit{E} (for example, in Al, Cu, Fe, Ag) have the same order of
magnitude, $\sigma _{s} /E \cong a_{0} /d_{c} $.

A contribution from ferroelastic fluctuations to the specific heat jump at
the superconducting transition temperature is also present in
low-temperature superconductors. In lead (Pb), the specific heat jump
$\Delta C_{P} $ [6] at the superconducting transition temperature $T_{c} $
is a sum of the Rutgers term and the term given by the equation (\ref{eq1}),

\begin{equation}
\label{eq3}
\Delta C_{P} = V\left( {T_{c} /4\pi}  \right)\left( {dH_{c} /dT}
\right)_{T_{c}} ^{2} + 0.01C_{P} .
\end{equation}

Here $H_{c} $ is the critical field (for type I superconductors) and
\textit{V} is the volume.

The equation (\ref{eq3}) is valid for diamagnetic metals (Pb, Sn, In, Tl). In
paramagnetic metals (Al, Ta), the specific heat jump $\Delta C_{P} $ at the
superconducting transition temperature $T_{c} $ is less than the value given
by the Rutgers formula, due to antiferromagnetic fluctuations in the
superconducting phase [5].

\begin{center}
---------------------------------------------------------------
\end{center}

[1] S.L.Bud'ko, N.Ni, and P.C.Canfield, Phys. Rev. B \textbf{79}, 220516 (R)
(2009).

[2] J.K.Dong, L.Ding, H.Wang, X.F.Wang, T.Wu, X.H.Chen, and S.Y.Li, New
J.Phys. \textbf{10}, 123031 (2008).

[3] T.Hirayama, M.Nakagawa, and Y.Oda, Solid State Commun. \textbf{113}, 121
(1999).

[4] F.V.Prigara, arXiv:0811.1131 (2008).

[5] F.V.Prigara, arXiv:0708.1230 (2007).

[6] G.S.Zhdanov, \textit{Solid State Physics} (Moscow University Press,
Moscow, 1961) [G.S.Zhdanov, \textit{Crystal Physics} (Oliver and Boyd,
Edinburgh, 1965)].

\end{document}